\documentclass[useAMS,usenatbib,fleqn]{mn2e}
\usepackage{graphicx}
\usepackage{amssymb}
\newcommand\msun{{\,M_\odot}}

\newcommand\zsun{{\rm \,Z_\odot}}
\newcommand{\unit}[1]{\ensuremath{\, \mathrm{#1}}}
\usepackage{aecompl}
\usepackage[T1]{fontenc}
\usepackage{threeparttable}
\usepackage{caption}
\usepackage{times}

\newcommand{\cMpc}{~\mbox{comoving}~\mbox{Mpc}}

\newcommand{\ckpc}{~\mbox{comoving}~\mbox{kpc}}

\newcommand{\cmci}{~\mbox{cm}^{-3}}

\newcommand{\K}{~\mbox{K}}

\title[Recovery from population~III supernova explosions]{Recovery from population~III supernova explosions and the onset of
second generation star formation}
\author[Jeon et al.]{Myoungwon Jeon$^{1}$\thanks{E-mail:
myjeon@astro.as.utexas.edu}, Andreas H. Pawlik$^{2}$, Volker Bromm$^{1}$ and Milo\v s Milosavljevi\'{c}$^{1}$\\
$^{1}$Department of Astronomy, University of Texas, Austin, TX 78712, USA \\
$^{2}$Max-Planck-Institut f\"ur Astrophysik, Karl-Schwarzschild-Strasse 1, 85748 Garching bei M\"unchen, Germany}

\begin{document}

\date{}


\maketitle
\topmargin-1cm

\label{firstpage}

\begin{abstract}
We use cosmological simulations to assess how the explosion of the
first stars in supernovae (SNe) influences early cosmic
history. Specifically, we investigate the impact by SNe on the host
systems for Population~III (Pop~III) star formation and explore its
dependence on halo environment and Pop~III progenitor mass. We then
trace the evolution of the enriched gas until conditions are met to
trigger second-generation star formation. To this extent, we quantify
the recovery timescale, which measures the time delay between a
Pop~III SN explosion and the appearance of cold, dense gas, out of
which second-generation stars can form. We find that this timescale is
highly sensitive to the Pop~III progenitor mass, and less so to the
halo environment. For Pop~III progenitor masses, $M_{\ast}=15\msun$, 
$25\msun$, and $40\msun$ in a halo of $5\times10^5\msun$, recovery times are 
$\sim10$\,Myr, $25$\,Myr, and $90$\,Myr, respectively. For more massive
progenitors, including those exploding in pair instability SNe,
second-generation star formation is delayed significantly, for up to a 
Hubble time. The dependence of the recovery time on the mass of
the SN progenitor is mainly due to the ionizing impact of the
progenitor star. Photoionization heating increases the gas pressure
and initiates a hydrodynamical response that reduces the central gas
density, an effect that is stronger in more massive. The gas around lower mass Pop~III stars remains 
denser and hence the SN remnants cool more rapidly,
facilitating the subsequent re-condensation of the gas and 
formation of a second generation of stars. In most cases, the
second-generation stars are already metal-enriched to $\sim2-5\times10^{-4}\zsun$, 
thus belonging to
Population~II. The recovery timescale is a key quantity governing the
nature of the first galaxies, able to host low-mass, long-lived
stellar systems. These in turn are the target of future deep-field
campaigns with the {\it James Webb Space Telescope}.
\end{abstract}


\section{Introduction}
\label{Sec:Intro}

One of the most intriguing aspects of modern cosmology is the
formation of the first stars and galaxies. The first stars, so-called
Population~III (Pop~III), were responsible for the transition of the
Universe from its simple, metal-free state to one of ever increasing
complexity. Therefore, exploring their feedback on the pristine
intergalactic medium (IGM) has been one of the key themes in
elucidating early cosmic evolution
\citep[e.g.,][]{Loeb2013,Wiklind2013}. Pop~III stars started forming
at redshifts as early as $z\gtrsim30$ out of primordial gas in dark
matter (DM) minihaloes of total mass $M\sim10^5-10^6\msun$
(e.g. \citealp{Haiman1996}; \citealp{Tegmark1997};
\citealp{Bromm2002}; \citealp{Yoshida2003}).  Once the first stars
began emitting ionizing radiation, the gas inside minihalo hosts was
photoheated to a few $10^4$\,K.  As a consequence, pressure-driven
outflows significantly suppressed central gas densities. After their
brief lifetime of $\sim10^6-10^7$\,yr, Pop~III stars may have exploded
as supernovae (SNe) or collapsed directly in black holes (BHs), possibly 
without explosion. The subsequent evolution of the surrounding IGM was 
shaped by the feedback from such Pop~III remnants (e.g. \citealp{Barkana2001};
\citealp{Bromm2004}; \citealp{Ciardi2005}; \citealp{Bromm2009}).
\par
If a Pop~III star died in a SN explosion, the ambient gas, already
rarified by the radiation-hydrodynamical reaction to the star's
ionizing radiation, was shock heated and hydrodynamically evacuated
from the minihalo as the shock traversed the halo. The dispersal of
heavy elements and dust from the first SNe initiated the long history
of cosmic chemical enrichment (e.g. \citealp{Karlsson2013}). This
enrichment was critical for star formation because the metal-enriched
dusty gas was able to cool to the temperature floor set by the cosmic
microwave background, $T_{\rm CMB}=$\,2.7\,K\,$ (1+z)$, similar to the
characteristic temperatures of star-forming clouds in the nearby
Universe. Such low temperatures were hard to reach in metal-free
primordial gas, in which molecular hydrogen was the main available
coolant. Therefore, the initial stellar mass function (IMF) is
expected to have changed from the top-heavy IMF, predicted for
Pop~III, to the low-mass dominated IMF of Population~II (Pop~II), once
enrichment levels exceeded a critical threshold.  Whether this
threshold was linked to fine-structure line cooling or dust-continuum
cooling is still debated, but there is general agreement that such a
`critical metallicity' played an important role
(e.g. \citealp{Omukai2000}; \citealp{Bromm2001b};
\citealp{Schneider2002}; \citealp{BrommLoeb2003}; 
\citealp{Omukai2005}; \citealp{Schneider2010}; \citealp{Schneider2012};
 \citealp{Chiaki2014}; \citealp{Ji2014}). If, on the other hand, a Pop~III star ended its
life as a BH via direct collapse, subsequent gas accretion onto the
remnant may have resulted in a miniquasar---a stellar mass black hole
accreting near the Eddington limit ---that
emitted X-rays (e.g., \citealp{Glover2003}; \citealp{Kuhlen2005};
\citealp{Milos2009a,Milos2009b}; \citealp{Alvarez2009};
\citealp{Park2011}; \citealp{Venkatesan2011}; \citealp{Wheeler2011};
\citealp{Jeon2012,Jeon2014}; \citealp{Xu2014}).
\par
The fate of a Pop~III star was determined by its mass and degree of rotation.
Non-rotating Pop~III stars with masses between $10\msun$ and $40\msun$ are 
thought to end their lives as core collapse SNe, or as highly 
energetic pair-instability supernovae (PISNe) for masses in the range of $140\msun \lesssim
M_{\ast} \lesssim 260\msun$ (\citealp{Heger2002}), or directly collapse into a BH for other 
progenitor masses (\citealp{Heger2003}). Stellar rotation may introduce significant changes
to this picture (\citealp{Maeder2012}). Core-collapse of 
rapidly rotating stars may trigger energetic hypernovae 
(e.g. \citealp{Umeda2005}), and the lower mass-limit for a PISN could be
reduced to $\sim 85\msun$
(\citealp{Chatzopoulos2012,Yoon2012}).
\par
The characteristic mass scale of Pop~III stars remains an open question 
(e.g. \citealp{Bromm2013}). Previously, the first stars were thought to be 
isolated and rather massive, with a characteristic mass of $\sim100\msun$ 
(e.g., \citealp{Abel2002}; \citealp{Bromm2002}; \citealp{Yoshida2006}). 
However, recent improved simulations have identified a ubiquitous disk 
fragmentation mode, revising the typical mass scale downwards to a few 
10$\msun$ \citep{Turk2009,Stacy2010,Clark2011a,Prieto2011,Smith2011, 
Greif2011,Greif2012,Stacy2012,Dopcke2013}. Moreover, radiative feedback has recently 
been shown to put an additional limit to the masses of these stars (e.g., \citealp{Hosokawa2011}; 
\citealp{SGB2012}; \citealp{Hirano2014}; \citealp{Susa2014}). Consequently, Pop~III stars were more likely to end their lives as core 
collapse SNe, rather than PISNe. In addition, observed abundance patterns 
in the atmospheres of metal-poor Galactic halo stars, expected to form 
out of gas enriched by Pop~III stars, have until recently not exhibited the strong
elemental odd-even effect and the absence of any neutron-capture elements, 
which would be manifest signatures for PISNe (\citealp{Aoki2014}). Therefore, very massive stars were 
rare, making them hard to detect, in agreement with some theoretical predictions 
(e.g., \citealp{Salvadori2007}; \citealp{Karlsson2008}), or Pop~III stellar masses
may have been limited to $\lesssim140\msun$ (Tumlinson 2006; 
\citealp{Heger2010}; \citealp{Joggerst2010}; \citealp{Keller2014}), or the paucity of chemical evidence for 
PISNe is due to the fact that the elements synthesized by PISNe were widely dispersed due to the PISN's large explosion energies (\citealp{Cooke2014}).
\par

\begin{table*}
\centering
\begin{threeparttable}[b]
\caption[Table]{Properties of host haloes, Pop~III progenitors, and SNe.}
\begin{tabular}{c c c c c c c c c}
\hline\hline
Halo & $M_{\rm vir}$ [$\msun$] & $z_{\rm form}$ & $M_{\ast}$ [$\msun$] & $t_{\rm life}$ [Myr] & $\dot{N}_{\rm ion}$ [$\rm s^{-1}$] & $E_{\rm SN}$ [ergs] & $y$ & Recovery time [Myr] \\
\hline
\textsc{halo1} & $5\times 10^5$&28&15 & 10 & $1.86\times 10^{48} $ &$1.0\times10^{51} $& 0.05 & 8.9\\
\textsc{halo1} & $5\times 10^5$&28&25 & 6.4 & $7.58\times 10^{48} $ & $1.0\times 10^{51}$& 0.05 & 24\\
\textsc{halo1} & $5\times 10^5$&28&40 & 3.8 & $2.47\times 10^{49}$ & $0.6, 1.0\times 10^{51}$& 0.05 & 92, 92\\
\textsc{halo1} & $5\times 10^5$&28&200 & 2.2 & $2.69\times 10^{50}$ & $10^{52}$& 0.5 & 300\\
\textsc{halo2} & $9\times 10^5$&25&25 & 6.4 & $7.58\times 10^{48} $ & $1.0\times 10^{51}$ & 0.05 & 3.7\\
\textsc{halo2} & $9\times 10^5$&25&40 & 3.8 & $2.47\times 10^{49}$ & $1.0\times 10^{51}$ & 0.05 & 6\\
\textsc{halo3} & $3\times 10^5$&23&40 & 3.8 & $2.47\times 10^{49}$ & $1.0\times 10^{51}$ & 0.05 & 140\\
\hline
\\
\end{tabular}\par
\begin{tablenotes}
\item[] From left to right, the columns show the host minihalo identifier, the total virial mass of the 
minihalo $M_{\rm vir}$, the Pop~III star's formation redshift $z_{\rm
  form}$ and mass $M_\ast$, the main-sequence
lifetime $t_{\rm life}$, the ionizing photon emission rate ${\dot N}_{\rm ion}$ (\citealp{Schaerer2003}), the 
SN energy $E_{\rm SN}$ and metal yield fraction $y$ (Table~2 in \citealp{Karlsson2013}), and recovery time
for return of gas into the center of the host halo (see Section
\ref{Sec:Results}). The recovery time for the SN explosion of energy $10^{52}$ ergs is taken from \citet{Greif2010}, who 
used identical initial conditions.
\end{tablenotes}
\label{table:SNs}
\end{threeparttable}
\end{table*}

There has been a number of numerical studies investigating the impact of 
early SNe on the gas surrounding the explosion sites and on the assembly 
process of the first galaxies (\citealp{Bromm2003}; \citealp{Kitayama2005}; 
\citealp{Greif2007,Greif2010}; \citealp{Whalen2008};
\citealp{Wise2008}; \citealp{Maio2010}; \citealp{BlandHawthorn2011};
\citealp{Vasiliev2012}; \citealp{Ritter2012}; \citealp{Wise2012}; \citealp{Chen2014}). 
Since the first stars were initially estimated to have very large masses $M_{\ast}\gtrsim100\msun$, most 
of these early studies focused on the impact of the more energetic PISNe. For instance, \citet{Greif2010} 
and \citet{Wise2008} followed the impact of a single PISN with an
explosion energy of $10^{52}$\,ergs until an atomic-cooling halo assembled at $z\approx10$. They found that even a single PISN explosion could enrich the region out of which the first galaxies formed to supercritical metallicities $Z > 10^{-3} \zsun$, such that
the first galaxies would typically host already metal-enriched Pop~II stellar systems.
\par
In contrast to PISNe, the impact of the less energetic core collapse SNe critically depended on the 
Pop~III progenitor mass and their environments, as was already found in one-dimensional, 
radiation-hydrodynamic simulations (\citealp{Kitayama2005}; \citealp{Whalen2008}; \citealp{Chiaki2013}). 
For instance, the gas in minihaloes with masses close to the cooling threshold for collapse, 
$\sim10^5-10^6\msun$, was more vulnerable to  
photoionization heating by the stellar progenitor, which decreased the density of the gas into which the SN 
exploded. Such conditions allowed SN ejecta to more easily escape the haloes. On the
other hand, the ionizing radiation from Pop~III stars 
inside more massive $\sim 10^7\msun$ host systems remained more strongly confined, such that the SN remnants
encounter mostly intact high-density clouds. Most of the SN explosion energy then was rapidly radiated away, 
and this impeded the escape of the metal-enriched gas to the IGM. 
\par
\citet{Ritter2012} investigated 
such a case of prompt fallback in a three dimensional, cosmological
grid-based hydrodynamic simulation, where a 40$\msun$ SN exploded in a
minihalo of $\sim10^6\msun$. 
The SN remnant remained partially trapped within the
minihalo. Roughly half of the ejecta turned around and fell back toward
the center of the halo, with $10\%$ of the ejecta reaching the center
in $\sim 10\,\textrm{Myr}$. The average metallicity of the combined
returning ejecta and the pristine cosmic web filaments feeding into the halo
center was $\sim0.001-0.01\,Z_\odot$, but the two fluids
remained unmixed until accreting onto a central hydrostatic core,
unresolved in the simulation.  
Ritter et al.\ concluded that if Pop~III stars had moderate
masses, they could
promptly enrich the host minihalos with metals and trigger Pop~II star formation.
\par
While these pioneering studies brought into focus the key physics, they either considered 
only a few specific cases, not exploring the dependence on a
varying cosmic environment, or employed somewhat 
simplified initial conditions. Taking into account realistic cosmological initial conditions 
and exploring a broader parameter space is clearly desirable. As an example, an 
insufficient number of ionizing photons from $10-40\msun$ Pop~III stars to achieve
substantial gas evacuation from the host system
might produce a highly asymmetric pre-explosion density distribution. 
The ensuing SN blastwave 
would then preferentially propagate along low-density channels, usually perpendicular to the 
cosmic filaments, resulting in highly non-uniform metal dispersal (e.g., \citealp{Wise2008}; 
\citealp{Greif2010}).

\par
To address the environmental 
complexity of early SN feedback, we here report on a suite of cosmological radiation-hydrodynamics simulations.
We followed the evolution of the SN-processed gas to the point at which the second generation of stars began
to form out of this metal-enriched material.
We specifically ask: {\it What is the delay between the initial Pop~III star formation event
and the re-establishment of conditions for subsequent star formation?}
Addressing this question is crucial for understanding the assembly 
process of the first galaxies and their properties
(\citealp{Bromm2011}). 
\par
The severity
of the Pop~III feedback is reflected in the mass scale of the first galaxies, since
stronger feedback requires the assembly of more massive host systems for the
SN-processed gas to be able to cool and recondense. This mass scale for the
first galaxies in turn determines their luminosities and number density, with
important implications for early reionization
(for reviews see, e.g., \citealp{Barkana2001}; \citealp{Barkana2009}; \citealp{Robertson2010}).
Pushing the high redshift frontier by observing the first galaxies is one of the main 
goals of next-generation facilities, such as the {\it James Webb Space Telescope} (JWST), the 
planned extremely large ground-based telescopes, the Atacama Large Millimitre Array (ALMA), 
and the Square Kilometre Array (SKA). The observability of the first galaxies will be determined 
by their star formation efficiencies and stellar IMFs (e.g., \citealp{Johnson2009}; \citealp{PMB2011};
\citealp{Zackrisson2011}).
\par
Recently, it has been suggested that the strength of SN driven
outflows may be responsible for the observed dichotomy between carbon-enhanced and
carbon-normal metal-poor stars \citep{Cooke2014}. These authors 
invoke two classes of SNe with greatly different times for recovery
from supernova explosions, such that
rapid recovery is connected to weak explosions with large carbon
overabundances, 
and slow recovery with strong explosions with normal carbon yields.
Pinning down gas dynamics in the aftermath of Pop III supernovae is
clearly essential for completing the picture of how the early
Universe evolved and what kind of long-lived relics it left in the nearby Universe. 
\citet{Ritter2014} suggested another explanation for carbon-enhanced metal-poor stars. Because the entropy to which the reverse shock raises the SN ejecta decreases with the ejecta mass coordinate, the central, iron-bearing ejecta are raised to a higher entropy and predominantly convect outward without cooling.  This prevents the bulk of the Pop III iron yield from returning to the halo center to enrich the second-generation star forming event.  Having lower entropy, the carbon-bearing ejecta are more susceptible to cooling and can return to the central cloud, enriching new stars.
\par
This paper is organized as follows. Our numerical methodology is
described in Section \ref{Sec:Metho} 
and the simulation results are presented in Section \ref{Sec:Results}. Finally, our
main findings are summarized in Section \ref{Sec:SC}. For consistency, all distances are expressed in physical (proper) units
unless noted otherwise.

\begin{figure}
\begin{center}
  \includegraphics[width=0.55\textwidth]{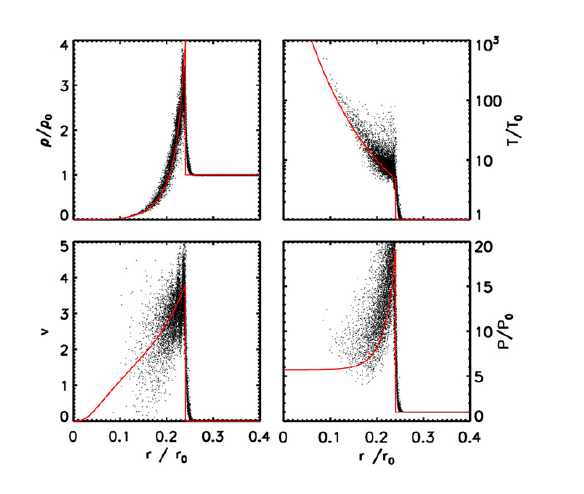}
\end{center}
    \caption{Test of SN blast wave dynamics in the absence of
      radiative cooling. 
    The shocked gas properties are shown at $t_{\rm sh}=0.02$ after the
    SN explosion with $E_0=1$. The red solid line is the self-similar Sedov-Taylor 
    solution. The agreement between the simulation and the self-similar solution is
    reasonable and the location of the shock is captured very well. 
    \label{Fig:Sedov}}
\end{figure}

\section{Numerical methodology}
\label{Sec:Metho}

\subsection{Gravity, hydrodynamics, and chemistry}
We carry out our investigation using the $N$-body/TreePM Smoothed Particle 
Hydrodynamics (SPH) code \textsc{gadget} (\citealp{Springel2005}; \citealp{Springel2001};
our specific implementation is based on that discussed in \citealp{Schaye2010}). 
Our simulations start with a snapshot from the earlier zoomed simulation of
\citet{Greif2010}, which was initialized at $z=99$ in a periodic box of size $1 \cMpc$. 
The initial conditions in \citet{Greif2010} were generated by assuming a $\Lambda$CDM cosmology with a matter 
density parameter of $\Omega_{\rm m}=1-\Omega_{\Lambda}=0.3$, 
baryon density $\Omega_{\rm b}=0.04$, present-day Hubble expansion rate $H_0
= 70\unit{km\, s^{-1} Mpc^{-1}}$, spectral index $n_{\rm s}=1.0$, and
normalization $\sigma_8=0.9$. We follow the evolution until the gas affected 
by the radiation of the Pop~III progenitor star and the ensuing SN explosion has recollapsed, 
reaching conditions where a second round of star formation can take place.
\par
\begin{figure*}
\begin{center}
  \includegraphics[width=0.75\textwidth]{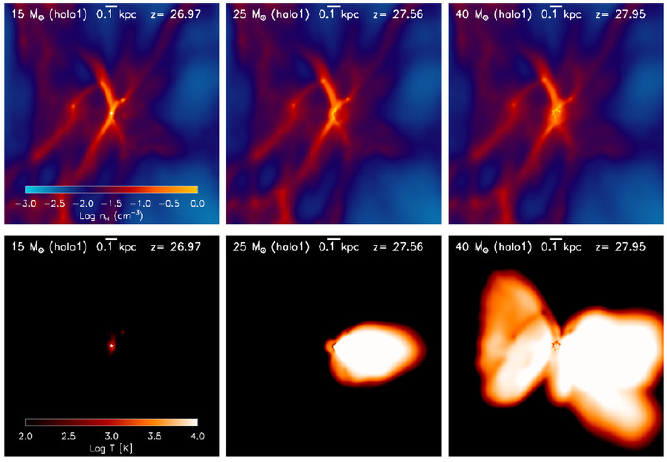}
\end{center}
  \caption{Projected gas density (top) and temperature (bottom) in a slab of size 35 kpc 
  (comoving) and thickness 7 kpc (comoving), centered on the locations of the 
  $15\msun$, $25\msun$, and $40\msun$ Pop~III stars (from left to right) in \textsc{halo1} 
    at the end of each star's lifetime and just before the subsequent explosion 
    in a SN. The ionizing photons from the stars preferentially 
    propagate into the voids, 
  perpendicular to the high density cosmic filaments, forming noticeable H~II regions around the 
  $25\msun$ and $40\msun$ stars. The HII region around the $15\msun$ star is unable to 
  outrun the hydrodynamical shock it generates, leaving it 
  in a highly compact state. \label{Fig:HIIs}}
\end{figure*}

Employing nested refinement of the cosmological initial conditions, 
the masses of dark matter (DM) and 
SPH gas particles in the highest resolution region with an approximate linear size of $300 
\ckpc$ are $m_{\rm DM} \approx 33\msun$ and $m_{\rm SPH} \approx 5 \msun$, respectively. 
Therefore, the baryonic mass resolution is $M_{\rm res} \equiv N_{\rm 
ngb} m_{\rm SPH}\approx 240\msun$, where $N_{\rm ngb} = 48$ is the number 
of particles in the SPH smoothing kernel \citep{Bate1997}. This 
resolution makes it possible to follow the gas evolution to a density 
of $n_{\rm H}\simeq 10^4 \unit{cm^{-3}}$ characteristic of the quasi-hydrostatic loitering state
before the onset of the runaway collapse that ultimately leads to Pop~III star formation
(\citealp{Abel2002}; \citealp{Bromm2002}). We adopt a Plummer-equivalent gravitational 
softening length of 70 comoving pc for both dark matter and baryonic particles. 
\par
We use the primordial chemistry and cooling network in Greif et al. (2010), where
all relevant cooling mechanisms are taken into account, specifically, H and He collisional ionization, 
excitation and recombination cooling, bremsstrahlung, inverse Compton cooling, and collisional 
excitation cooling of $\rm H_2$ and HD. The code self-consistently solves the non-equilibrium chemistry 
for 9 species ($\rm H, H^{+}, H^{-}, H_{2}, H^{+}_2 , He, He^{+}, He^{++},$ and $\rm e^{-}$), as well as for the three 
deuterium species $\rm D, D^{+}$, and HD. Moreover, after the SN explodes, the primordial gas is polluted 
with heavy elements. The code allows us to follow the chemical evolution of C, O, and Si, ejected by the SN 
explosion, assuming solar relative abundances, and includes the corresponding
metal line collisional excitation and cooling rates (Glover \& Jappsen 2007). 
The chemical network comprises the key species, 
$\rm C, C^{+}, O, O^{+}, Si, Si^{+}$, and $\rm Si^{++}$.
\par

\begin{figure}
\begin{center}
  \includegraphics[width=0.5\textwidth]{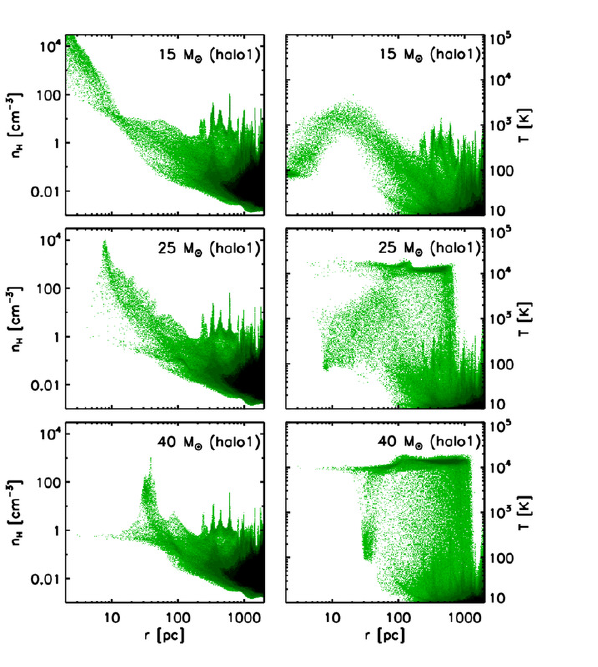}
\end{center}
      \caption{Density (left) and temperature (right) profiles at the end of the progenitor's life, altered by the radiation from 
      the $15\msun$, $25\msun$, and $40\msun$ Pop~III stars (from top to bottom) in \textsc{halo1}. The asymmetric 
      structure of the H~II regions around the $25\msun$ and $40\msun$
      stars, shown in Fig.~\ref{Fig:HIIs}, is reflected in the
      bifurcation of the gas properties inside $r\lesssim
      50\,\textrm{pc}$: the gas lying along the cosmic filaments remains dense 
      $n_{\rm H}\gtrsim10\unit{cm^{-3}}$ and cold $T<10^3$\,K, while that in the less dense regions is photoevaporated to 
      $n_{\rm H}\lesssim1\unit{cm^{-3}}$ and photoheated to $T\approx10^4$\,K. The H~II region created by the 
      $15 \msun$ star remains too small to significantly impact the distribution of the gas in the halo.   \label{Fig:scatter_HIIs}}
\end{figure}

\subsection{Pop~III properties and ionizing radiative transfer}
The characteristic mass of the first stars is a matter of intense
research. Recent high-resolution simulations have revised the
characteristic scale downward to tens of solar masses and have
suggested a broad Pop~III IMF, as explained in the Introduction. In this work, we aim to explore a
representative subset of progenitor masses. We provide the properties
of Pop~III progenitors and SN explosions in Table~\ref{table:SNs}
(\citealp{Heger2002}; \citealp{Schaerer2003};
\citealp{Karlsson2013}). To consider environmental effects on the
evolution of Pop~III SN remnants, we insert Pop~III stars in three
different minihaloes inside the most highly resolved
region. Properties of these halos are provided in
Table~\ref{table:SNs}.
\par
We start our simulations at $z \approx 30$, corresponding to the time
just before the first Pop~III star has formed. Once the hydrogen
number density exceeds $n_{\rm H, max}=10^4 \cmci$, the
highest-density SPH particle is converted into a collisionless sink
particle, subsequently accreting neighboring gas particles until its
mass has reached the target Pop~III progenitor star mass (see
Table~\ref{table:SNs}). 
\par
The sink particle then becomes a source of ionizing radiation, emitted
at a rate averaged over the main sequence lifetime computed by
\citet{Schaerer2002}.  The radiation spectrum is assumed to be a
black-body spectrum at temperature $10^{4.76}\K$, $10^{4.85} \K$, and
$10^{4.9}\K$ for the stars of 15, 25, and 40 solar mass, respectively 
(\citealp{Schaerer2002}). We transport this ionizing radiation using
the radiative transfer (RT) code \textsc{traphic}
\citep{Pawlik2008,Pawlik2011}. \textsc{traphic} solves the
time-dependent RT equation in SPH simulations by tracing photon
packets emitted by source particles in a photon-conserving manner.
The RT is controlled by a set of parameters, which we choose identical to
those employed in \citet{Jeon2014}, except that we here use two frequency
bins with bounding energies of [13.6 eV, 400 eV, 10 keV]. Recombination radiation is
treated in the Case B approximation (for a discussion of the
applicability of this approximation see, e.g., \citealp{R2014}, and
references therein).
\par
In addition to the
ionizing radiation from Pop~III stars, we also track the Lyman-Werner
(LW) radiation that dissociates $\rm H_2$ and HD. The latter is treated
in the optical thin limit with a self-shielding
correction. For further details on our RT methodology and the treatment of 
LW radiation we refer the reader to \citet{Jeon2014}.
\subsection{Supernova energy injection}
\label{Sec:SNEnIn}
We explode each SN at the end of the progenitor's main-sequence
lifetime. The explosion energy is distributed among the nearest
$\bar{N}_{\rm ngb}=32$ neighbor SPH particles with equal weight. This
establishes a strong pressure jump relative to the surrounding medium,
in turn launching a blast wave.
\par
Within SPH schemes, particles are often assigned individual time
steps, sorted in a hierarchical fashion, such that only a fraction of
particles is `active' at any time. Only the properties of active
particles are updated at any given time.  Therefore, if a SN blast
wave alters the internal energies of neighboring gas particles, only
the active particles among them are able to react to this sudden
change. Inactive particles cannot be updated right away and this
introduces serious errors into the hydrodynamics. These errors can be
reduced by imposing a sufficiently small upper limit on the individual
time steps. However, using such a global constraint on the time steps 
is computationally expensive. 
\par
Instead, in order to elicit 
a prompt response to energy injection, we employ the time
step limiter originally suggested by \citet{Saitoh2009} and developed
further by \citet{Durier2012}. This limiter restricts the time steps
of neighboring particles, specifically those within an SPH smoothing
kernel, such that the time step ratios do not exceed a given factor
$f_{\rm step}$. However, \citet{Durier2012} have shown that when only the time step
limiter is used, energy conservation may be strongly violated if energy is 
injected at a random time. To tackle
this problem, \citet{Durier2012} implemented an update scheme where at
the time of the SN explosion {\it all} $\bar{N}_{\rm ngb}$ neighboring
particles inside of the explosion volume become active. In our 
simulations, we use both the time step limiter
and the time step update by adopting the value of $f_{\rm step}=4$. This 
ensures the accurate integration of the SN explosion, and is substantially less computationally expensive
than comparably accurate simulations that impose a small global upper limit
on the particle timestep (e.g., \citealp{Greif2007}). For reference, the typical timesteps in our 
simulations are $\Delta t\approx 10^3-10^4$\,yr. Those can be reduced 
by an order of magnitude when needed, and the change in timesteps is controlled both by the timestep limiter and update procedure.
\par

We tested our implementation by simulating the self-similar Sedov-Taylor (ST) problem, 
which has a known analytic solution. Fig.~\ref{Fig:Sedov} shows the
shocked gas properties affected by a SN explosion of energy $E_0=1$, at a time 
$t_{\rm  sh}=0.02$ after energy injection. Initially, $N=128^3$ gas particles
were uniformly distributed in a glass-like configuration 
with mean density $\rho_0=1$ (\citealp{White1996}). The initial internal energy was set to
$u_0=10^{-3}$.  We show the self-similar ST solution with red
solid lines (e.g., \citealp{Shu1992}, \citealp{Durier2012}). The location of the shock front,
$R_{\rm sh}\sim(E_0 t_{\rm sh}^2/\rho_0)^{1/5}$, where $t_{\rm sh}$ is
the evolution time and $\rho_0$ the density of the surrounding medium,
is consistent with the numerical result. Directly behind the shock,
the gas accurately approximates the Rankine-Hugoniot jump conditions
for the density, velocity, and pressure, [$\rho_{\rm sh}, v_{\rm sh},
  P_{\rm sh}]$ $\equiv$ [$4\rho_0$, $3\dot{r}_{\rm sh}/4$, $3\rho_0
  \dot{r}_{\rm sh}^2/4$] $\approx[4, 3.78, 19]$. Here, $\dot{r}_{\rm
  sh}\approx5$ is the shock velocity.

\begin{figure*}
\begin{center}
  \includegraphics[width=0.75\textwidth]{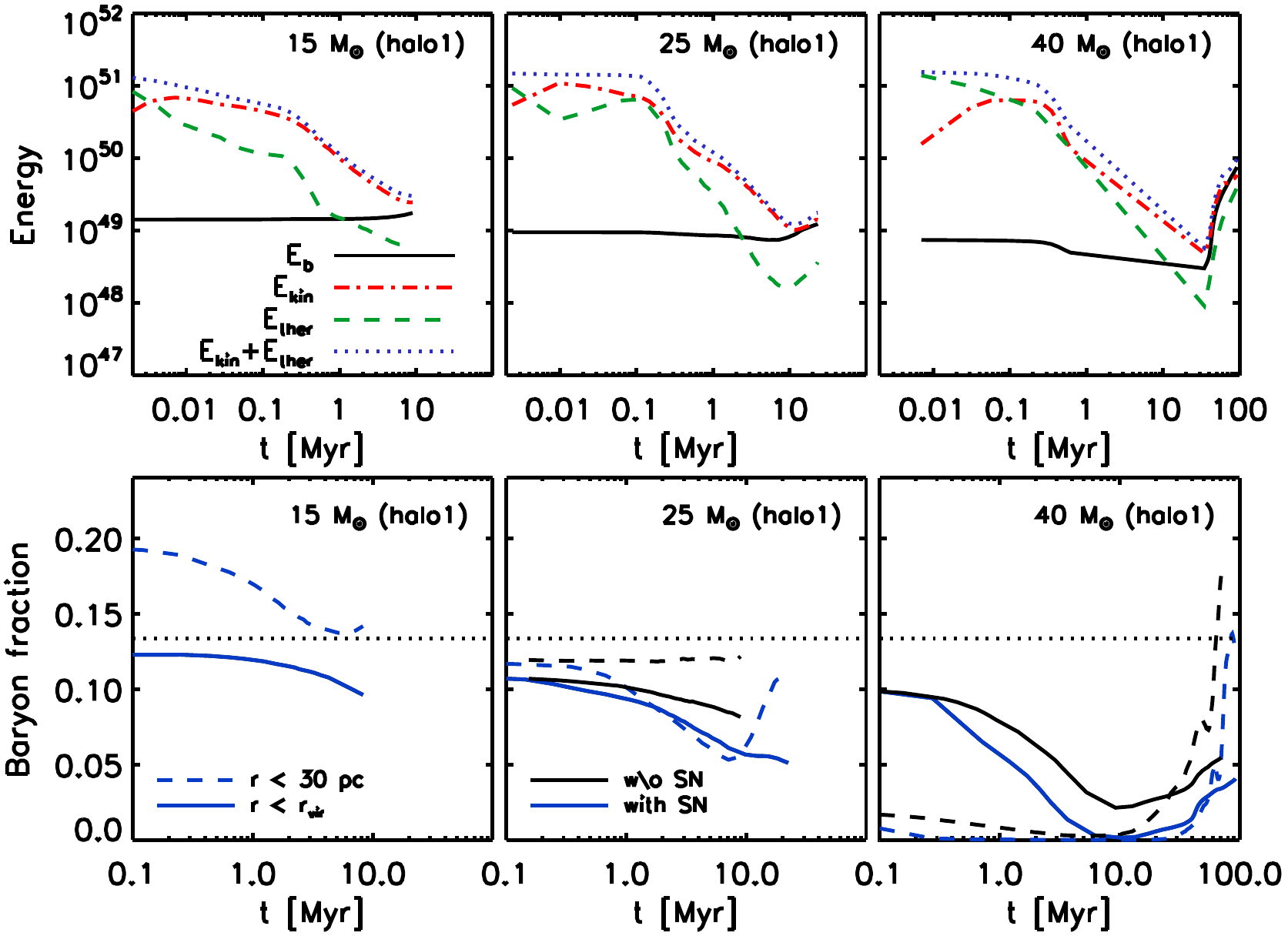}
\end{center}
  \caption{The evolution of the binding (solid), kinetic (dot-dashed), and thermal (dashed) energy of the gas particles 
  inside the virial radius of the host halo, up to the point at which the central gas density reaches 
  $n_{\rm H}=10^4\unit{cm^{-3}}$ (top panels). The horizontal axis
  shows the time elapsed since the SN explosion. The bottom panels 
  show the evolution of the baryon fraction computed within the central 30 pc (dashed) and within the virial radius (solid). 
  The blue lines show the baryon fraction affected by both the SN and its progenitor, while the black lines represent the 
  case when only the photoionization by the Pop~III star is taken
  into account. The horizontal dotted lines show the universal baryon
  fraction.
   \label{Fig:energy}}
\end{figure*}

\subsection{Metal transport}
\label{Sec:Metal}
The modeling of the chemical transport in SPH
simulations is a difficult task. In pure SPH schemes, there is no mass flux 
between particles.  Also, for SPH to resolve the turbulent cascade
that is the pre-condition for mixing would require much higher resolution
than attainable in cosmological simulations. 
Here, we implement a subgrid prescription for compositional diffusion 
as in \citet{Greif2009a}, where metal 
transport between particles was modelled as a diffusion process (\citealp{Klessen2003}). 
The discretized diffusion equation in the SPH formalism is 
\begin{equation}
\label{eq:diffusion}
\frac{dc_i}{dt} = \sum_j {K_{ij}(c_i - c_j)},
\end{equation}
with
\begin{equation}
K_{ij} = \frac{m_j}{\rho_i \rho_j} \frac{4D_i D_j}{(D_i + D_j)} \frac{r_{ij} \cdot \nabla_i { W_{ij}}}{r_{ij}^2},
\end{equation}
where $c_i$ is the concentration of SPH particle $i$, $m_i$ the
particle mass, $\rho_i$ the density at the particle location $\mathbf{r}_i$, $W_{ij}$ the smoothing kernel
evaluated at 
$r_{ij}=|\mathbf{r}_i-\mathbf{r}_j|$, the separation between particles
$i$ and $j$, and $D_i$ is a diffusion coefficient associated with
particle $i$.

If a diffusion process is to meaningfully emulate turbulence-assisted
chemical mixing, the diffusion coefficient must somehow reflect the
amplitude of the local turbulent cascade on subgrid
scales. For example, a cascade may be expected in the presence of
strong shear that would excite
the Kelvin-Helmholtz instability.  More generally, one might
anticipate the strength of the subgrid turbulent cascade to scale with
the amplitude of gas
motions on the smallest resolved scales, ignoring the motions on
larger scales.  In grid-based hydrodynamic simulations, 
the loss of kinetic energy to subgrid scales has been
modeled with a diffusion term with diffusivity 
$\sim \Delta (\rho K_{\rm sgs})^{1/2}$, where $\Delta$ is the grid spacing
and $K_{\rm sgs}$ is the kinetic energy density on subgrid scales
\citep{Schmidt2006,Schmidt2011}.  In SPH, we replace the grid spacing
with the SPH smoothing length $\tilde{l}$ and the kinetic energy
density with $\rho\tilde{v}^2$, where $\tilde{v}$ is the velocity
dispersion within the SPH kernel, which for particle $i$ equals
\begin{equation}
\tilde{v}_i^2 = \frac{1}{N_{\rm ngb}} \sum_{j=1}^{N_{\rm ngb}} |\mathbf{v}_i-\mathbf{v}_j|^2,
\end{equation}
where $\mathbf{v}_i$ is the velocity of particle $i$.
The diffusion coefficient is then proportional to $\rho \tilde{v} \tilde{l}$ and we adopt
\begin{equation}
D=2 \rho \tilde{v} \tilde{l}.
\end{equation}
A direct integration of Equation (\ref{eq:diffusion}) then yields
\begin{equation}
\label{eq:concentration_advanced}
c_i (t_0 + \Delta t) = c_i(t_0) e^{A\Delta t} + \frac{B}{A} (1-e^{A\Delta t}),
\end{equation}
where $A=\sum_j {K_{ij}}$ and $B=\sum_j {K_{ij} c_j}$.  We use
Equation (\ref{eq:concentration_advanced}) to advance the
concentrations of SPH active particles.  In evaluating
the right hand side, the densities and smoothing lengths are evaluated
at the advanced time $t_0+\Delta t$.
\par
Initially, the metals from a SN are evenly distributed among the
$\bar{N}_{\rm ngb}$ neighboring particles. Therefore, the initial 
metallicity of neighboring SPH particles is defined by 
\begin{equation}
Z_i = \frac{m_{{\rm metal}, i}}{m_{\rm SPH} + m_{{\rm metal}, i}},
\end{equation}
where the metal mass of individual SPH particles is $m_{\rm metal, i}=
y M_{\ast}/\bar{N}_{\rm ngb}$, where $y$ 
is the metal yield as given in Table~\ref{table:SNs}. 
\par
\begin{figure*}
\begin{center}
  \includegraphics[width=0.75\textwidth]{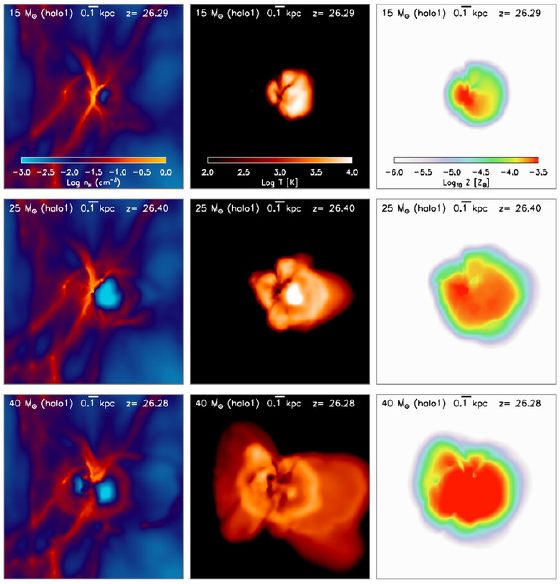}
\end{center}
    \caption{Projected hydrogen number densities (left), temperatures
      (middle), and metallicities (right) in the same slab as in
      Fig.~\ref{Fig:HIIs} for Pop~III progenitor star masses 15$\msun$, 25$\msun$, 
    and 40 $\msun$, respectively (from top to bottom). The snapshots
    show the time at which the gas blown out by photoionization and the
    SN blastwave 
begins to flow back into the minihalo. This corresponds to $\approx5$\,Myr, $\approx9$\,Myr, and $\approx15$\,Myr 
    after the SN explosion for the 15$\msun$, 25$\msun$, and 40 $\msun$ stars, respectively. We emphasize that the simulations 
    are identical except for the progenitor's ionizing luminosities, hence any differences in the results are entirely due
    to the sensitivity of the dynamics of the SN heated and accelerated gas to the pre-processing by ionizing radiation from the progenitor. \label{Fig:snaps}}
 \end{figure*}
 
\begin{figure}
\begin{center}
  \includegraphics[width=0.45\textwidth]{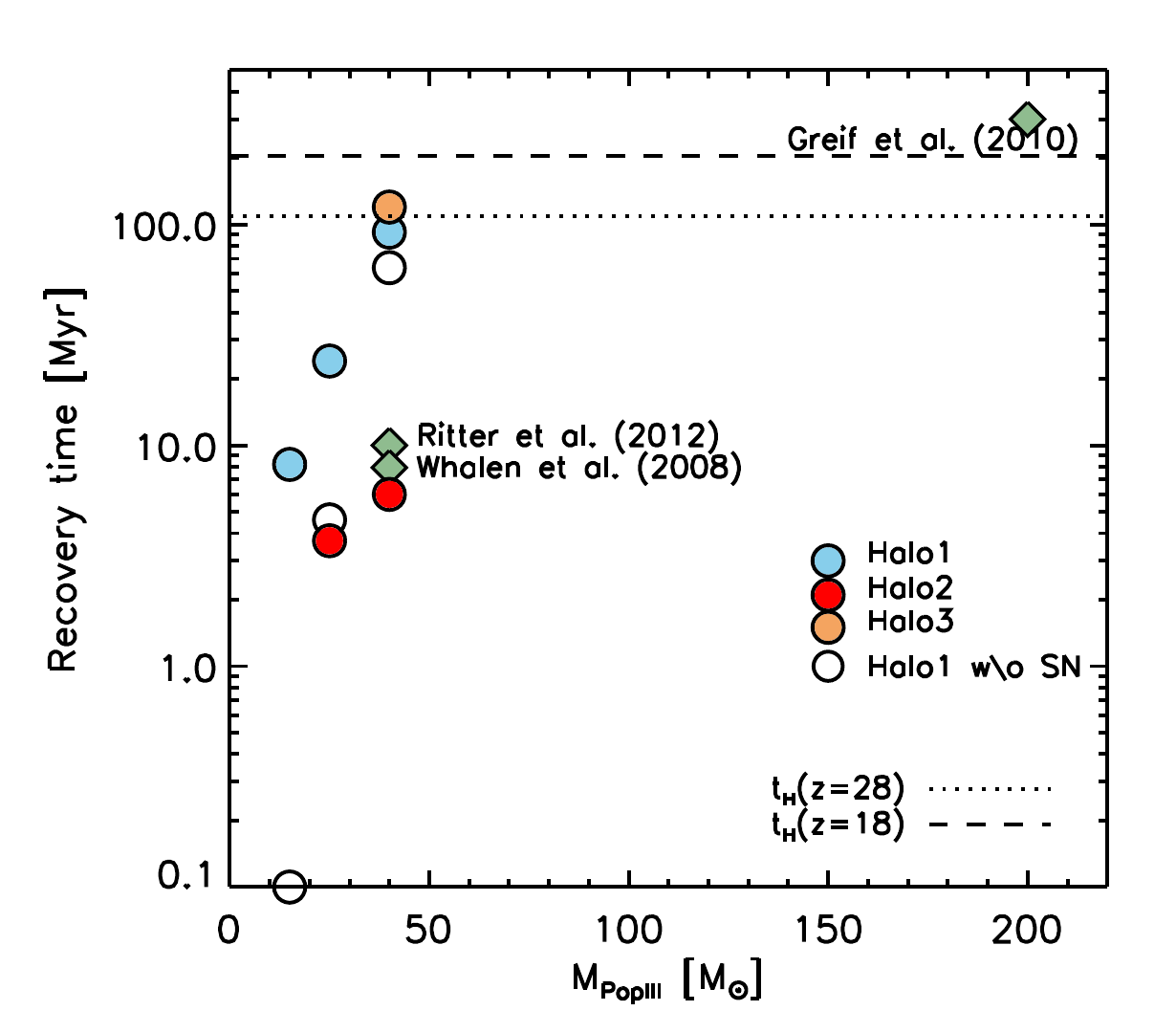}
\end{center}
    \caption{Recovery times for the central gas disturbed by
      photoionization and the SN blastwave 
    as a function of Pop~III progenitor mass. Colors are used to distinguish
    between haloes. The recovery times from simulations of stars in \textsc{halo1} with an H~II region but no SN 
    in which gas outflow was driven solely by the radiation-hydrodynamical response to
    photoionization heating, are marked 
    by open circles. For comparison, results from previous studies are shown by diamonds.
    \label{Fig:SN}}
\end{figure} 
\section{Results}
\label{Sec:Results}

In this section, we present our results. First, in
Section~\ref{Sec:HII}, we show how the initial configuration for the
SN explosion is affected by photoionization by the Pop~III progenitor
star as a function of the star's mass, and how it is affected by the
environment provided by the host halo. In Section~\ref{Sec:SN}, we
describe the evolution of the SN blast wave. In Section~\ref{Sec:SSF},
we quantify the Pop~III star's destructive impact in terms of the time
scale on which baryons return to the center of the halo after the SN
explosion, establishing conditions for the formation of a second
generation of stars.  In addition, we also assess the properties of
this second generation event, in particular the metallicity of the
star-forming gas.

\subsection{Photoionization}
\label{Sec:HII}
In our simulation box, the first star forms at $z\approx28$ out of
primordial gas inside a $M_{\rm vir}\approx5\times10^5\msun$
minihalo. Fig.~\ref{Fig:HIIs} shows the
projected densities and temperatures of the gas surrounding the
Pop~III stars with masses $15\msun$, $25\msun$, and $40\msun$ (from
left to right). The sizes and shapes of the H~II regions, as depicted in
Fig.~\ref{Fig:HIIs}, reflect the detailed transport of the
ionizing photons emitted by the stars in the inhomogeneous density field of a given halo.
Some gas around the 25$\msun$ and 40$\msun$ stars is photoheated to
temperatures $\sim10^4$\,K, forming an extended H~II region, whereas
the ionized region created around the $15\msun$ star remains trapped 
inside the hydrodynamic shock, giving rise to a D-type
ionization front. In the latter case, the time for ionizing photons to
break out of the shock front, $t_{\rm B}$, is longer than the stellar
lifetime, $t_{\rm B}>t_{\ast}$. Our results are consistent with those in
\citet{Alvarez2006} and \citet{Whalen2008}, where the authors find
that no ionizing radiation will escape into the IGM for
$M_{\ast}<15\msun$.  Specifically, in our simulations, the breakout
times are 3.5\,Myr and 1.95\,Myr for the 25$\msun$ and 40$\msun$ stars
(in \textsc{halo1}). 
\par
The high ionizing luminosities of
very massive Pop~III stars with $M_{\ast}\gtrsim100\msun$ are sufficient to
photoevaporate even the densest gas in the halo, producing a smooth
H~II region (e.g., \citealp{Greif2009}; \citealp{Jeon2012}). On the other hand, 
as also noted by \citet{Abel2007},
\citet{BlandHawthorn2011}, and \citet{Ritter2012}, at lower ionizing
luminosities (say, from Pop~III stars in the mass range
$10\msun\lesssim M_{\ast} \lesssim40\msun$), the densest clouds in the
halo, often aligned with the filaments of the cosmic web, may
self-shield and remain intact. The resistance to photoevaporation at
low ionizing luminosities is also expected from analytical
considerations \citep{Bertoldi1990}.  The residual neutral clouds cast
shadows inside the halo and produce a highly asymmetrical, sometimes
butterfly-shaped H~II region.

These asymmetric structures are evident in Fig.~\ref{Fig:scatter_HIIs}
in the form of two distinct branches in the density and temperature radial
profiles. Except for the $15\msun$ star, there are photo-heated gas
particles with $n_{\rm H}\lesssim 1\unit{cm^{-3}}$ and
$T\approx10^4$\,K near the star, while at the same time, the gas
embedded in a neighboring cosmic filament remains dense ($n_{\rm
  H}\gtrsim100\unit{cm^{-3}}$) and cold ($T\sim200$\,K). The sizes of
the H~II regions are $\approx 800$\,pc and $\approx 1$\,kpc in
diameter for the $25\msun$ and $40\msun$ stars, respectively (in
\textsc{halo1}).  We note that even for a fixed Pop~III stellar mass,
the size of the resulting H~II region could vary by a factor of 2,
depending on halo environment. For instance, the H~II regions of a
$40\msun$ Pop~III star in \textsc{halo2} and \textsc{halo3} reach
$\approx1.8$ kpc and $\approx2$ kpc, respectively. Allowing for a solitary Pop~III star to form in a star-forming cloud is an approximation that focuses on 
  the feedback by an individual Pop~III star. Following the evolution of the cloud up to high 
  densities ($n_{\rm H}\gtrsim 10^4 \unit{cm^{-3}}$), sufficient to see further fragmentation, is beyond the scope of this work. None of the haloes hosting 
  Pop~III stars considered in this work show other baryon density peaks high enough to meet the criterion for star formation 
  within their virial radii.
  \par
\subsection{Supernova blastwaves}
\label{Sec:SN}

The SN is triggered at the end of the progenitor's lifetime, when the preceding 
photoionization has significantly modified the conditions in the ambient gas.
Thus, the explosions occur in a wide range of environments,
as discussed in Section~\ref{Sec:HII}.
In most simulations, we choose an explosion energy of
$E_{\rm SN}=10^{51}$\,ergs. We also perform an additional run for the $40\msun$ star (in \textsc{halo1}) with
$E_{\rm SN}=6\times 10^{50}$\,ergs. Due to the limited resolution, the
explosions are initiated after the end of the 
free expansion phase, where the ejecta mass roughly equals the swept-up mass.
The time when the free expansion phase ends (\citealp{Kitayama2005}) 
\begin{eqnarray}
t_{\rm fe}&=&1.6\times10^4\unit{yr}  \nonumber \\ 
&\times&\left(\frac{n_{\rm H}}{1\,\unit{cm^{-3}}}\right)^{-1/3}\left(\frac{M_{\rm ej}}{200\msun}\right)^{5/6}\left(\frac{E_{\rm
  SN}}{10^{51}\unit{ergs}}\right)^{-1/2}
\end{eqnarray}
is $\approx 200$\,yr, $\approx 2.8\times10^3$\,yr, and $\approx 4.7\times10^3$\,yr for the 15$\msun$, 25$\msun$, and 40$\msun$
stars, respectively. The latter two expand freely on time scale $t_{\rm fe}$ comparable to
the time step, $\Delta t\approx 10^3-10^4$\,yr, encountered in
our simulations. The injection of the SN energy leads to a sudden increase in the mean temperature of the
particles near the explosion to $T_{\rm init}\approx 5\times 10^7$\,K. 

Subsequently, the thermal SN energy is rapidly converted into
kinetic energy (see the top panels of Fig.~\ref{Fig:energy}), at $\approx2-3\times
10^{-3}$\,Myr for the $15\msun$ and $25\msun$ cases, and at $\approx$ 0.1\,Myr for the $40\msun$ case.
As the resulting blast wave propagates outward, the gas cools by adiabatic expansion. Afterwards, forming a dense shell,
the energy of the gas behind the shock is radiatively lost primarily via inverse Compton scattering
against the CMB, and also by free-free emission. In particular for the $15\msun$ and $25\msun$ stars, due
to the initial configuration where the density of nearby gas remains high ($n_{\rm H}\gtrsim10^3 \unit{cm^{-3}}$),
the free-free cooling, operating on a timescale, $t_{\rm cool, ff}\propto 1/n_{\rm H}$, plays an important role in decreasing temperatures to $10^6$\,K within
$<0.5$\,Myr. Once the gas cools to $\sim10^6$\,K, H and He resonance processes begin to dominate the cooling.
As the blastwave energy dissipates, it eventually drops below the total binding energy of the gas,
$E_{\rm B}=3GM_{\rm vir}M_{\rm b}/(5R_{\rm vir})$, where $M_{\rm
  b}\simeq \Omega_{\rm b}/\Omega_{\rm m}\times M_{\rm vir}$ is the
total baryonic mass, and $M_{\rm vir}$ and $R_{\rm vir}$ are the
virial mass
and radius of the given halo, respectively.
\par
In Fig.~\ref{Fig:energy} (bottom panels), we illustrate the evolution
of the baryon fraction, both within the virial radius of the host halo
and within the central 30\,pc from the SN explosion site, up to the
point when the central gas becomes dense enough to support new star
formation. For comparison, we also show the evolution of the baryon
fraction when only photoionization by the Pop~III star was enabled,
with no subsequent SN explosion (denoted by the black lines in
Fig.~\ref{Fig:energy}).  Fig.~\ref{Fig:energy} shows that
photoionization by the 40$\msun$ Pop~III star was already strong
enough to blow out most of the gas, at $r\lesssim 30$\,pc, resulting
in a significant decrease of the baryon fraction to $f_{\rm
  b}\approx0.01$ prior to the SN explosion. At 9\,Myr after the death
of the star, about 80$\%$ of the gas mass is driven out of the virial
radius.  The resulting deficit of gas at the center lasts for 42\,Myr
and 3.5\,Myr with and without SN explosion, respectively. The relative
impact of the SN explosion on the central region is stronger in the
$25\msun$ case. This can be seen by comparing the baryon fraction with
and without SN explosion, the latter remaining nearly constant over
the run time, and dropping by $40\%$ at $\approx9$\,Myr when the SN is
included. After $\approx8$\,Myr, $\approx9$\,Myr, and $\approx50$\,Myr
for the $15\msun$, $25\msun$, and $40\msun$ stars (in \textsc{halo1}),
the central baryon fraction starts to increase as the gravitational
force pulls back the gas. The haloes then attain central densities of
up to $n_{\rm H}=10^4\unit{cm^{-3}}$ within the next 2\,Myr, 14\,Myr,
and 40\,Myr, respectively.

\begin{figure*}
\begin{center}
  \includegraphics[width=0.75\textwidth]{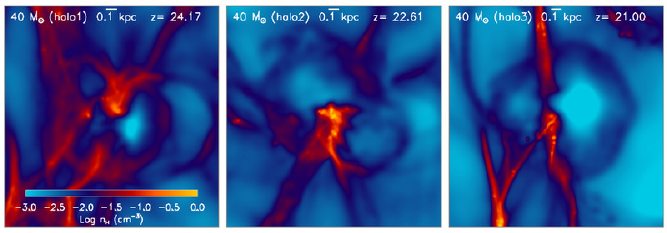}
\end{center}
      \caption{Hydrogen number densities 25\,Myr after the SN explosion for the $40\msun$ 
      progenitor in different environments. The central gas in \textsc{halo2} (middle) has already recovered 
      to form a second generation of stars, while that in \textsc{halo1} (left) and \textsc{halo3} (right), 
      which are less massive, still shows 
      low central densities. \label{Fig:Haloes}}
\end{figure*}

In Fig.~\ref{Fig:snaps}, we present the projected hydrogen number densities, temperatures, and metallicities for the 15$\msun$, 
25$\msun$, and 40$\msun$ stars (from top to bottom). The snapshots show the time at which the gas flow reverses from net outflow to inflow.
After 1\,Myr, for the $40\msun$ star, the forward shock passes $r_{\rm sh}\approx120$ pc, beyond 
the virial radius of the host halo, $\rm R_{\rm vir}\approx75$ pc. As the shock propagates, the central gas densities 
decrease over $1\,\textrm{Myr}$ from $n_{\rm H}\approx0.8-1$ $\rm cm^{-3}$ to $n_{\rm H}\approx0.01$ $\rm cm^{-3}$ for the 25$\msun$ 
and 40$\msun$ stars, and from $n_{\rm H}\approx10^4$ $\rm cm^{-3}$ to a few $n_{\rm H}\approx$ $10^3$ 
$\rm cm^{-3}$ for the $15\msun$ star. Similar to the expansion
direction of the H~II region, the SN energy is preferentially transferred into underdense regions, perpendicular to 
the direction of cosmic filaments.  

\subsection{Onset of second generation star formation}
\label{Sec:SSF}
\subsubsection{Recovery timescale}
We next estimate the recovery timescale, defined as the time delay
between the first star formation event in a given halo and the
recollapse of the gas affected by photoionization and the SN blastwave
to high densities, such that a second generation of stars can
form. This timescale sensitively depends on the SN progenitor mass. If
we consider the same SN input energy, the difference in recovery
timescale arises solely from the pre-explosion configuration shaped by
the stellar-mass-dependent ionizing luminosity. Specifically, more
massive Pop~III stars have a stronger impact on the ambient gas,
suppressing central gas densities prior to the explosion more
strongly.  As a consequence, the SN remnants suffer smaller
radiative losses, and are thus able to escape further into the IGM. In
our simulations, we find recovery times of $\approx8$\,Myr,
$\approx24$\,Myr, and $\approx90$\,Myr for the $15\msun$, $25\msun$,
and $40\msun$ stars, respectively (in \textsc{halo1}).  These
timescales are shorter by 8\,Myr, 15\,Myr, and 30\,Myr when no SN
explosion is included (see open circles in Fig.~\ref{Fig:SN}). We emphasize 
that since we use a fixed SN energy, $E_{\rm SN}=10^{51}\unit{ergs}$, for 
all three progenitors, the difference in the recovery time arises from differences in the photoionization feedback during the pre-explosion phase.

\par
In Fig.~\ref{Fig:SN} we also mark the results from \citet{Greif2010}, where they followed the
evolution of the gas after a $200\msun$ PISN explosion. Given that we here use the same initial
conditions as in \citet{Greif2010}, a direct comparison is possible. The recovery time in their work was of order the Hubble time,
$\approx300$\,Myr, at $z\approx10$. On the other hand, the work by \citet{Ritter2012},
where they studied the gas evolution and metal enrichment from a $40\msun$ core collapse
SN with $E_{\rm SN}=10^{51}\unit{ergs}$, suggests a shorter recovery timescale, $\lesssim10$\,Myr.
The reason for the much shorter recovery timescale compared to our simulation of a SN explosion with 
a $40 \msun$ progenitor, is twofold. First, their halo mass at recovery was $2\times10^6 \msun$, which is four
times higher than the mass of our \textsc{halo1}. Second, as they explain in their
Section 2.4, their treatment of the pre-explosion
photoionization by the Pop~III progenitor artifically underestimated
the degree of photoheating, such that
gas densities remained high. Any radiative losses from the SN blastwave thus were
overestimated. In addition to the core-collapse SNe, we show the case of a
$40\msun$ hypernova with an energy
$3\times10^{52}\unit{ergs}$ exploding in a substantially more massive halo of $1.2\times10^7\msun$
(\citealp{Whalen2008}). Here, the gas again quickly
recollapses in $\approx8$\,Myr. This broad range of recovery timescales
suggests a strong dependence on environment.
\par
To explore this environmental dependence, we have performed additional
simulations by assuming that a $40\msun$ or a $25\msun$ Pop~III star
forms in \textsc{halo2} with mass $\approx10^6\msun$ and
\textsc{halo3} with mass $\approx3\times10^5\msun$. We use the same
explosion energy $E_{\rm SN}=10^{51}\unit{erg}$ in all cases. Due to
the deeper potential well of \textsc{halo2}, the deleterious impact of
the Pop~III stars is weaker than in \textsc{halo1} and
\textsc{halo3}. In \textsc{halo2}, a larger fraction of the gas is more
likely to resist photoevaporation by the $40\msun$
Pop~III star, maintaining central densities of $n_{\rm
  H}\approx10^4\unit{cm^{-3}}$. These are substantially higher than
the typical densities in \textsc{halo1} when exposed to the same
Pop~III stellar radiation.  Consequently, the SN remnants encounter
higher density gas, thus experiencing more rapid energy dissipation
through radiative cooling, leading to a shorter recovery
time. Specifically, the estimated recovery time for the $40\msun$ star
in \textsc{halo2} is $\approx6$\,Myr, similar to the recovery time found in 
\citet{Ritter2012}, which investigated a $40\msun$ Pop~III star
exploding in a halo with mass $\approx10^6\msun$.
\par
On the other hand, the impact of the Pop~III feedback in
\textsc{halo3} extends to much longer times: it takes more than
100\,Myr for the gas expelled by photoionization and the SN blastwave
of a $40\msun$ star to recollapse to high densities. This is largely
due to the rather isolated environment of \textsc{halo3} which
impedes gas accretion onto the halo. In Fig.~\ref{Fig:Haloes}, we
compare the density field in different haloes 25\,Myr after the SN
explosion for the $40\msun$ case. In all haloes, the hydrodynamic
shock driven by the explosion propagates to $\approx400-600$\,pc. 
While the central gas in \textsc{halo2} (middle panel of Fig.~\ref{Fig:Haloes}) has already 
reached densities high enough to fulfill the condition for star formation,
densities in \textsc{halo1} and \textsc{halo3} are still low.

\begin{figure}
\begin{center}
  \includegraphics[width=0.45\textwidth]{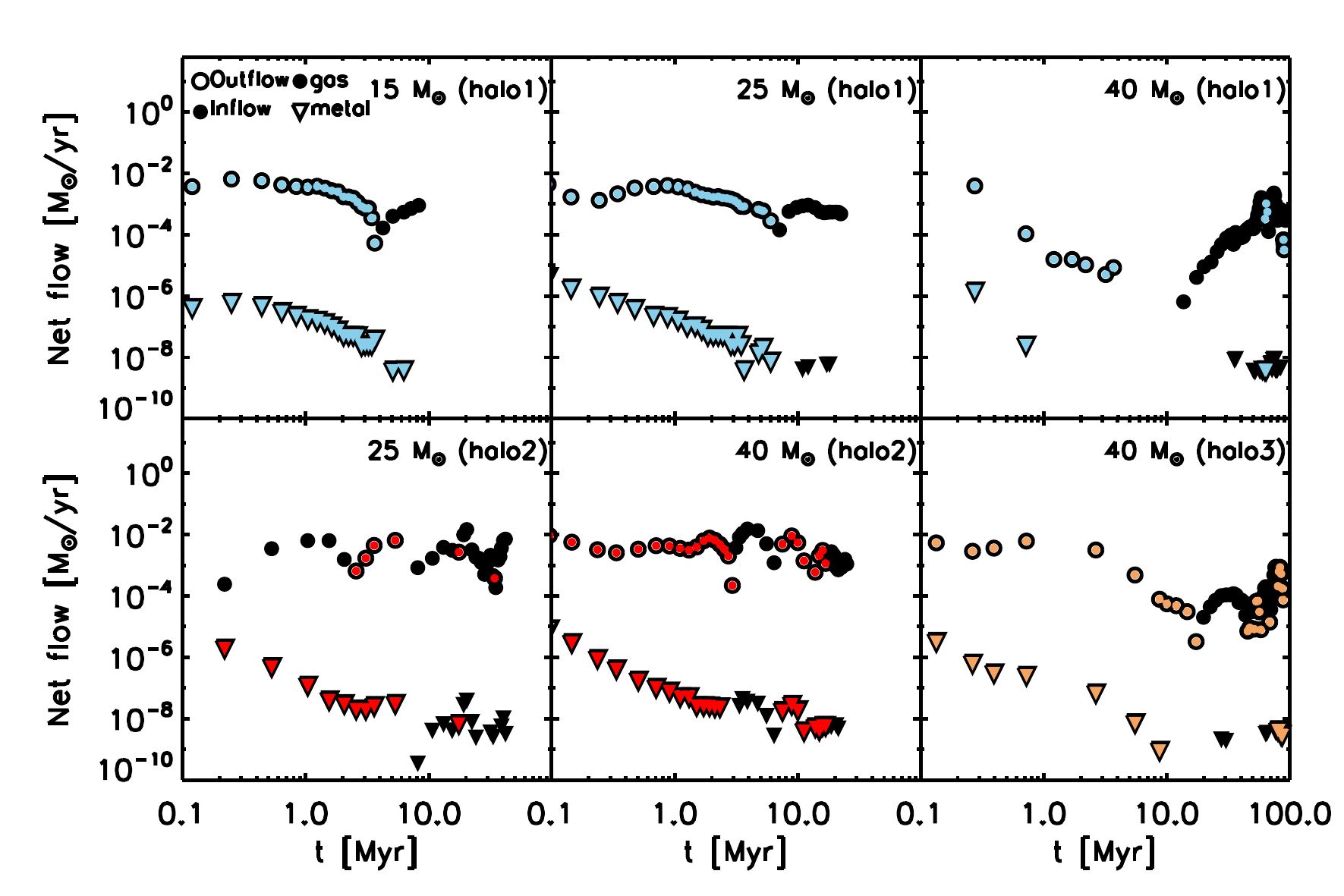}
\end{center}
  \caption{Gas (circles) and metal (upside down triangles) net mass flow rate through
  a sphere of radius 30 pc centered on the location of the SN explosion. Black and colored 
  circles distinguish between inflow and outflow. \label{Fig:Netflow}}
\end{figure}

Next, we address the question of when the gas flow changes sign,
reverting from net outflow to inflow. Fig.~\ref{Fig:Netflow} presents
the evolution of the gas (circles) and metal (triangles) flow through
a sphere of radius 30\,pc, centered on the location of the SN
explosion. Inflow dominates after $\approx6$\,Myr, $\approx8$\,Myr,
and $\approx11$\,Myr after the SN for the $15\msun$, $25\msun$, and
$40\msun$ stars (in \textsc{halo1}).  Along with the gas,
metal-enriched material is brought back in. The transition from net
outflow to inflow for the metals occurs a few million years after it
has occurred for the gas regardless of the metal content. This is because 
metal-free gas can also be delivered onto the central region along the filaments of the cosmic web. The
situation is more complicated in \textsc{halo2}. For the $40\msun$
case, the transition happens at $\approx3$\,Myr, but again reverting
into an outflow for the next 25\,Myr.  This unstable flow is due to
the environmental complexity, leading to unsteady accretion. We also
note that the stellar feedback from the $25\msun$ Pop~III star in
\textsc{halo2} is initially not strong enough to reverse the cosmological 
inflow of gas, resulting in a delayed outflow until $\approx2$\,Myr after the SN
explosion.

\subsubsection{Metallicity of star-forming gas}

\begin{figure}
\begin{center}
  \includegraphics[width=0.45\textwidth]{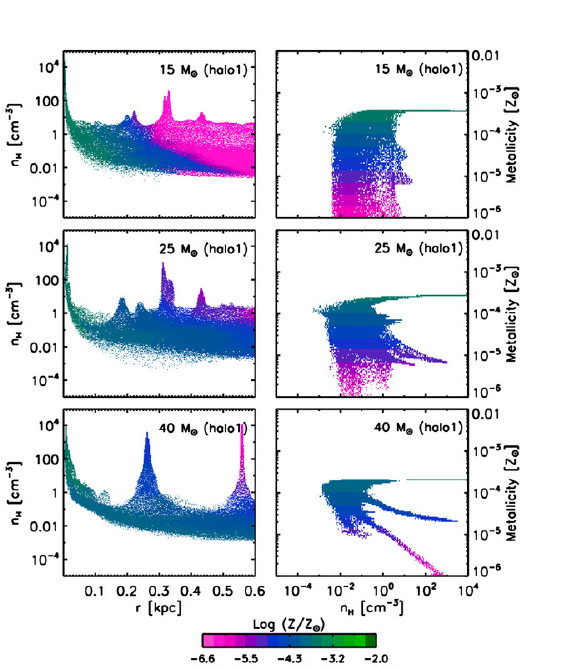}
\end{center}
   \caption{Hydrogen number density as a function of distance from the SN (left) and
   the metallicity as a function of the hydrogen number density
   (right) for the 15, 25, and
   40 $\msun$ Pop~III star (from top to bottom), all shown at the corresponding 
   recovery times.\label{Fig:Metals}}
\end{figure}

To what levels can the first SNe, triggered by Pop~III stars, enrich the surrounding
medium prior to the onset of second generation
star formation? For the $40\msun$ star, 
the total mass in heavy elements, $y M_{\ast}\simeq 2\msun$, is initially
distributed evenly among $\bar{N}_{\rm ngb}=32$ nearby SPH particles, resulting in
a metallicity of
$y M_{\ast}/(m_{\rm SPH}\bar{N}_{\rm ngb})=0.0125$. Afterwards, metal dispersal is modeled as a diffusion
process, as described in Section~\ref{Sec:Metal}. As seen in the top panels of Fig.~\ref{Fig:Metals},
at the end of the simulation the metals from the $15\msun$ explosion can penetrate to $r\approx600$ pc,
mainly enriching the underdense regions ($n_{\rm H}\approx0.01\unit{cm^{-3}}$) to $\sim5\times10^{-4}\zsun$,
while the high density regions ($n_{\rm H}\gtrsim10\unit{cm^{-3}}$) located far ($r\gtrsim300$ pc) from the halo, hosting the $15\msun$ star, are kept primordial. About 17$\%$, 20$\%$, and 42$\%$
of the gas mass within 600\,pc is enriched above $10^{-4}\zsun$ for the $15\msun$, $25\msun$, and
$40\msun$ stars, respectively. The typical extent to which the metals are dispersed is
$\approx550$ pc, $\approx1$ kpc, and $\approx1.5$ kpc, for the $15\msun$, $25\msun$, and
$40\msun$ stars, respectively. As expected, metal dispersion preferentially proceeds into
 the low-density phases of the IGM.
\par
Interestingly, in \textsc{halo1} the central gas at $r\lesssim10$ pc, where densities
are high enough to form stars,
is enriched to $2-6\times10^{-4}\zsun$. This is above the critical metallicity,
$Z_{\rm crit}\simeq 10^{-3.5}\zsun$, for enabling low-mass star formation via metal or dust grain cooling.
As shown in the left-bottom panel
of Fig.~\ref{Fig:Metals}, there are other high density peaks close to $n_{\rm H}=10^4\unit{cm^{-3}}$
within the central 600 pc. The high density gas ($n_{\rm H}\gtrsim100\unit{cm^{-3}}$) near 
$r=560$ pc has not yet been contaminated with metals, while the surrounding, relatively
low density gas, is enriched to $\gtrsim10^{-5}\zsun$. The high-density peak has avoided contamination
because the gas collapsed into the halo prior to the arrival of the metals, and
the mixing of metals into such pre-condensed structures was inefficient and slow.
On the other hand, the density peak near $r\approx260$\,pc is enriched to $\gtrsim10^{-5}\zsun$,
demonstrating the large range of possibilities.

\par
Whether such levels of enrichment suffice
to achieve the transition from primordial, high-mass dominated, star formation to
a normal, low-mass dominated, (Pop~II) mode, depends on the detailed physics
behind $Z_{\rm crit}$.
In elucidating this physics, significant uncertainties remain. By way of
fine-structure line cooling, mainly involving C~II and O~I, metal-enriched gas is able to fragment to form a cluster of 
Pop~II stars \citep[e.g.,][]{Bromm2001b,SSC2014b,SSC2014a}. The minimum metallicity required to enable such widespread fragmentation is $Z_{\rm crit}\simeq10^{-3.5}\zsun$.
Although with fine-structure cooling, thermal Jeans masses are of order a few $10\msun$, it may still be 
possible to form low-mass stars ($M_{\ast}\lesssim0.8\msun$) if some of the fragments get dynamically ejected from 
the cluster, thus prematurely curtailing gas accretion (e.g. \citealp{Ji2014}). A second pathway toward low-mass 
star formation is enabled when the initial dust-to-gas mass ratio, where the dust grains are produced during the 
first SN explosions (\citealp{Schneider2006}), exceeds a critical level of $\it{D}_{\rm crit}=\rm [2.6-6.3]\times10^{-9}$ (\citealp{Schneider2012}; \citealp{Chiaki2014}). The critical metallicity in this scenario is a 
few orders of magnitude below that for the atomic line cooling scenario. Therefore, the density peak at $\approx 260$ pc, enriched to a few
$10^{-5}\zsun$, is not capable of forming Pop~II stars based on atomic
line cooling alone, but it satisfies the condition for fragmentation set by dust cooling.
\par
Do current observations, obtained by the stellar archeology of low-mass, metal-poor stars,
provide us with hints for this theoretical debate?
Most metal-poor stars with $\rm [Fe/H]<-4.0$
found in the Hamburg/ESO survey (\citealp{Frebel2006}; \citealp{Christlieb2008}) satisfy the condition for fine-structure cooling
(\citealp{Frebel2007}).
In addition, the recently discovered star SMSS 0313-6708 has no detected iron, with an
upper limit of $\rm [Fe/H]<-7.1$, but exhibits a high carbon abundance,
$\rm [C/H]=-2.6$ (\citealp{Keller2014}), thus again exceeding the fine-structure
threshold. However, there are counter-examples, specifically the most metal-poor
star found to date, in which {\it all} elements, including C and O, are underabundant
\citep{Caffau2011}. Here, dust cooling seems to be required \citep{Klessen2012, Schneider2012, Chiaki2014}.
The empirical picture is thus complex, with possible roles for both cooling channels \citep{Ji2014}.

\section{Summary and conclusions}
\label{Sec:SC}
We have performed a suite of cosmological three-dimensional simulations to investigate
the recovery from Pop~III core-collapse SNe with a mass range for the progenitor
stars of $M_{\ast}=15-40\msun$.
We have also investigated the chemical impact of the SNe on the next episode of star
formation. 
\par
We found that the radiative feedback from Pop~III stars plays an important
role in setting the initial configuration for the subsequent SN by reducing the gas density 
prior to the explosion.  Unlike the photoionization by massive Pop~III stars with
$M_{\ast}\sim100\msun$, the radiative impact from less massive Pop~III stars with
$M_{\ast}\sim15-40\msun$ is less strong in evacuating gas
out of the host halo. Therefore, the ejecta from the SNe are
more likely to encounter a high density ambient medium, in which the ejecta cool
more rapidly and may be re-accreted earlier. We find that the time 
it takes for the SN-processed gas to fall back into the halo that hosts a Pop~III star with a
moderate mass is of the order of a few 10\,Myr. We also
explore the effect of the environment by exploding SNe in different host haloes.
We find that the impact of a Pop~III SN explosion in a $M_{\rm vir}\sim10^5\msun$ halo is more
violent than in a $M_{\rm vir}\sim10^6\msun$ halo.

\begin{figure}
\begin{center}
  \includegraphics[width=0.45\textwidth]{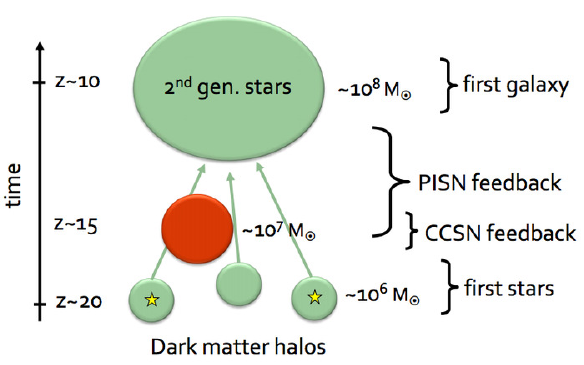}
\end{center}
    \caption{Assembly of a first galaxy under Pop~III feedback. Minihaloes that host Pop~III
    stars first collapse at $z\gtrsim30$ and then grow for $\sim300$\,Myr to assemble an atomically cooling galaxy in a
    $10^8\msun$ DM halo. The impact of a PISN
    is very destructive, delaying the gas collapse in the halo for $t_{\rm recover}\sim300$\,Myr
    (\citealp{Greif2010}). On the other hand, the delay of second generation star formation that is 
    introduced by Pop~III core-collapse SNe is shorter, of the order of a
    few 10\,Myr. The properties of the first stars, in particular their mass, thus govern the nature of the first galaxies. \label{Fig:FG}}
\end{figure}

In the context of hierarchical cosmology, structures assemble from the bottom up: smaller objects
collapse first, making bigger structures. The first galaxies, thus, are believed to have grown by
assembling roughly $\sim10$ Pop~III star-forming minihaloes
\citep[Fig.~\ref{Fig:FG} and ][]{Greif2008}. Theoretically, one often defines 
the first galaxies as the DM haloes with masses $M_{\rm vir}\sim10^8\msun$, collapsing at $z\simeq10$.
These objects are massive enough to retain the gas that was affected by previous star formation,
and this gas inside the galaxy is able to cool via atomic hydrogen
line emission. Previous works by, e.g., \citet{Wise2008} and \citet{Greif2010} 
show that a single PISN explosion in a minihalo can
be sufficiently violent, destroying the minihalo in which the PISN
explodes and enriching the central star-forming gas in the nascent galaxy to $10^{-3}\zsun$. 
\par
On the other hand, as shown in this work, the impact of 
core-collapse SNe with progenitor masses $15-40\msun$ on the delay of
gas collapse and thus on the next episode of star formation is prompt, 
$t_{\rm recover}\approx10$\,Myr. Depending on the halo environment and halo 
mass, this timescale may vary from 10\,Myr to 100\,Myr, but always remains substantially smaller 
than the delay of $t_{\rm recover}\sim300$\,Myr resulting from a PISN. 
This suggests that multiple star formation episodes, likely occurring in the Pop~II
mode, may be triggered at the site that was affected by
core-collapse Pop~III stars, even before the system evolves into a 
substantially more massive galaxy (e.g.,  \citealp{Wise2012}; \citealp{PMB2013}).
\par
Finally, we would like to point out that the range of primordial stellar masses remains uncertain. A recent investigation carried out 
by \citet{Hirano2014}, where they study $\sim$100 different cosmological sites of Pop~III star formation, shows that the masses 
of Pop~III stars might be from a few tens to hundreds of solar masses, depending mainly on gas accretion rates in 
the star-forming clouds. In addition, a recently observed metal-poor Milky Way star exhibits the elemental odd-even effect, possibly 
a chemical signature of PISNe (\citealp{Aoki2014}). Consequently, the relative importance of core-collapse SNe and PISNe in shaping 
the formation of the first galaxies will depend crucially on the Pop~III IMF.
\par
In summary, our simulations imply that the
central cold and dense gas can be enriched above $Z=10^{-4}\zsun$ by a single
core-collapse Pop~III SN. Given the expected multiple star
formation events before the collapse of the minihalo host into a $10^8\msun$ atomically cooling halo, it is likely 
that the central star-forming gas in the first galaxies is substantially polluted with metals, 
possibly initiating Pop~II star formation in the first galaxies.

\section*{acknowledgements}
We thank Raffaella Schneider for helpful comments. We are grateful to Volker Springel, Joop Schaye, and Claudio Dalla
Vecchia for letting us use their versions of \textsc{gadget} and their data
visualization and analysis tools. V.~B.\ and M.~M.\ acknowledge support
from NSF grant AST-1009928 and NASA ATFP grant NNX09AJ33G.
A.~H.~P.\ receives funding from the European Union's Seventh Framework
Programme (FP7/2007-2013) under grant agreement number
301096-proFeSsoR. The simulations were carried out at the Texas
Advanced Computing Center (TACC).

\end{document}